\documentclass[final,3p,times]{elsarticle}

\usepackage{amsmath}
\usepackage{amssymb}
\usepackage{amsthm}
\usepackage{natbib}
\usepackage{color}
\usepackage{graphicx}
\usepackage[normalem]{ulem}

\journal{Physics Letters A}

\begin{document}

\begin{frontmatter}


\title{Impurity induced Modulational instability in Bose-Einstein condensates}

\author{Ishfaq Ahmad Bhat, Bishwajyoti Dey}
\address{Department of Physics, Savitribai Phule Pune University, Pune 411007, Maharashtra, India
}
\begin{abstract}
By means of linear stability analysis (LSA) and direct numerical
simulations of the coupled Gross-Pitaevskii (GP) equations, 
we address the impurity induced modulational instability (MI) 
and the associated nonlinear dynamics in Bose-Einstein 
condensates (BECs).
We explore the dual role played by the impurities within the BECs 
\textemdash the instigation of MI and the dissipation of the 
initially generated  solitary waves. 
Because of the impurities, the repulsive BECs are even 
modulationally unstable and this tendency towards MI increases 
with increasing impurity fraction and superfluid-impurity coupling 
strength. However, the tendency of a given BEC towards the MI decreases 
with the decreasing mass of the impurity atoms while the 
sign of the superfluid-impurity interaction plays no role.
The above results are true even for 
attractive BECs except for a weak superfluid-impurity 
coupling, where the MI phenomenon is marginally suppressed 
by the presence of impurities. The dissipation of the 
solitons reduces their lifetime and is eminent for a larger 
impurity fraction and strong superfluid-impurity strength 
respectively. 

\end{abstract}

%\includegraphics[width=0.5\textwidth]{rho_1.0_rep_mis.pdf}
%\includegraphics[width=0.5\textwidth]{mi_pc_dgl1.png}

%\item We analyze the phenomenon of modulational 
%instability (MI) in a Bose-Einstein condensate (BEC) 
%coupled to a reservoir of uncondensed impurities.
%\item The presence of impurities in repulsive BECs makes them always modulationally unstable such that the tendency of a BEC towards MI increases with the increasing concentration of impurities.
%\item In repulsive BECs, the critical time for the development of MI decreases with the impurity fraction. On the other hand, the number of initially generated solitons increases with the impurity fraction.
%\item In attractive BECs, the critical  MI time either decreases or increases with the impurity fraction, depending on the strength of the superfluid-impurity coupling. For weak superfluid-impurity coupling, the MI time increases while if the superfluid-impurity coupling is much strong, the MI time decreases.
%\item The initially generated solitons undergo dissipation during their time evolution. The dissipation is large for a greater impurity fraction and superfluid-impurity coupling. Thereby reducing the lifetime of generated solitons.



\end{frontmatter}


%%%%%%%%%%%%%%%%% Introduction %%%%%%%%%%%%%%
\section{Introduction}\label{intro}
Modulational instability (MI) is the most prevalent 
phenomenon in nonlinear systems \cite{Zakharov_physicad_2009} 
wherein a continuous wave background becomes unstable 
against the growth of weak perturbations. As a result, localized 
solitary waves are formed due to the interplay between the 
intrinsic nonlinearity and dispersion. Although MI has been 
studied in nonlinear fibre optics \cite{Agrawal_nlo_2013}, 
fluid dynamics \cite{Brunetti_pla_2014}, plasma 
\cite{Taniuti_prl_1968}, and even scattering phenomena  
\cite{Malomed_pra_1993}, the MI in BECs has recently 
received great attention both theoretically \cite{Konotop_pra_2002, Theocharis_pra_2003,Kasamatsu_prl_2004,Li_pra_2005, Kasamatsu_pra_2006} 
and experimentally \cite{Everitt_pra_2017, Ferrier_pra_2018}. A variety 
of BEC applications, such as quantum phase transitions \cite{Kanamoto_pra_2003}, 
matter-wave amplification \cite{Inouye_nat_1999}, 
atom interferometry \cite{Anderson_pra_2003} and 
atom lasers \cite{Hagley_sci_1999} result from the efficient 
manipulation of the matter wave dynamics of an ultracold BEC 
by external fields. 
The MI in BECs and its associated dynamics is well described by 
the mean-field Gross-Pitaevskii (GP) equation which is nonlinear 
in nature. Scalar BECs are governed by the single component 
GP equations while the multicomponent BECs are described by coupled 
GP equations. The nonlinearity in a BEC is introduced via the 
\textit{s}-wave interactions that are manipulated precisely in 
experiments by the optical \cite{Fatemi_prl_2000,Blatt_prl_2011}
and magnetic \cite{Inouye_nat_1998, Cornish_prl_2000} Feshbach 
resonances. In addition to accounting for the MI in BECs, 
the GP equation exhibits different types of nonlinear 
structures, including solitons \cite{Strecker_nat_2002, Burger_prl_1999} 
and vortices \cite{Matthews_prl_1999, Abo_sci_2001}.
\par 
The presence of impurities in BECs offer a platform to investigate 
mass-imbalanced multicomponent systems \cite{Catani_pra_2012, Jorgensen_prl_2016, Hu_prl_2016, Javed_pra_2016}. 
The impurity-superfluid interactions in these 
imbalanced BEC systems realize Bose polarons which under strong 
interactions result into impurity-solitons \cite{Shadkhoo_prl_2015}.
The impurity-superfluid systems are also central towards the 
understanding of Kondo effect \cite{Hewson_kondo_1993}, 
spin transport \cite{Zvonarev_prl_2007},
spin-charge separation \cite{Kleine_prl_2008} and 
synthetic superfluid chemistry \cite{Edmonds_prr_2021}. 
While the number of impurities in a  BEC can be fixed 
either by the controlled doping \cite{Spethmann_prl_2012} 
or by transferring coherently a fraction of the condensate 
atoms from one hyperfine level to another by means of a radio-frequency (rf)
pulse \cite{Aycock_pnas_2017}.
\par 
The earlier works addressing MI in single component BECs \cite{Theocharis_pra_2003, Salasnich_prl_2003, Carr_prl_2004}
dictate that MI is a natural precursor to the formation 
of bright solitons. In a single component BEC, MI is possible only for 
a focusing (attractive) nonlinearity \cite{Higbie_prl_2002, Khawaja_prl_2002}. 
By tuning the atomic interactions in a 
BEC from repulsive to attractive, MI results in the formation of bright 
matter-wave solitons \cite{Strecker_nat_2002,Everitt_pra_2017,Nguyen_sci_2017}. 
In a binary BEC, MI is possible even for repulsive interactions 
whereby dark-bright solitary wave complexes are formed if the 
self-repulsion of each component is outbalanced by the 
cross-repulsion between the two components 
\cite{Goldstein_pra_1997, Kasamatsu_prl_2004,Kasamatsu_pra_2006}. 
The dark-bright soliton complexes are coherent 
structures consisting of a bright soliton in one component effectively trapped 
by a dark soliton in the second component. This results in the formation of domain walls
which realize the phase separation in the immiscible binary BEC \cite{Kasamatsu_prl_2004,Kasamatsu_pra_2006,Wen_pra_2012,Vidanovic_njp_2013}.
The MI phenomenon has also been studied extensively in BECs with synthetic 
gauge potentials \cite{Bhat_pra_2015,Otlaadisa_pra_2021}. 
In presence of gauge potentials, MI occurs even in the 
miscible binary BECs. Recently, the MI analysis has been extended to 
dipolar \cite{Ferrier_prl_2016, Schmitt_nat_2016} and 
density-dependent BECs \cite{Bhat_pre_2021}. In dipolar BECs, the MI results 
in the formation of quantum droplets while in density-dependent BECs 
unidirectionally moving chiral solitons are obtained. 
\par 
In the framework of the mean-field approach, this paper addresses 
the MI and the associated nonlinear dynamics of solitary wave formation 
in an effectively one-dimensional superfluid-impurity system. 
We consider a binary BEC system in which the condensate (superfluid) 
interacts with a dilute fraction (0-6\%) of impurity atoms. 
When the impurity fraction in a BEC exceeds 0.2\%, 
the impurity atoms are also Bose-condensed \cite{Aycock_pnas_2017}. 
The MI analysis in a binary BEC has usually been 
carried out for an equal distribution of atoms in the two components 
\cite{Kasamatsu_prl_2004,Kanamoto_pra_2003,Bhat_pra_2015}. 
However, by varying the impurity fraction and their interaction 
with the superfluid, we study the effect of mass imbalance on 
the MI phenomena in the BECs. Moreover, the impurity atoms can be either 
heavier or lighter or of the same mass as the superfluid atoms. A unit mass 
ratio can be achieved by transferring coherently a small fraction of 
superfluid atoms into another hyperfine state. Such a situation is neatly 
modelled by the experiments of Aycock \cite{Aycock_pnas_2017} and 
J{\o}rgensen \cite{Jorgensen_prl_2016}. By considering different 
superfluid-to-impurity atomic mass ratios, we additionally study 
its effect on the impurity-induced MI in BECs. 
\par 
The subsequent material is structured as follows. Section \ref{sec:model}
introduces the model and the corresponding coupled GP equations. 
The dispersion relation produced by the linear stability analysis (LSA)
is derived and discussed. 
Section \ref{sec:numerics} reports the results of numerical
simulations of the system under consideration. The paper is
concluded by Sec. \ref{sec:conc}.
%%%%%%%%%%% Model and MI analysis %%%%%%%%%%%%%%
\section{Model and MI Analysis}\label{sec:model}
We consider a one-dimensional (1D) BEC  with atomic 
mass $m_{1}$ and consisting of $N_{1}$ atoms in a state 
$\psi_{1}$ collisionally coupled to $N_{2}$ Bose-condensed 
impurity atoms of mass $m_{2}$ in the state $\psi_{2}$.
Such a projected BEC system is described by the following
coupled GP equations \cite{Edmonds_prr_2021, Aycock_pnas_2017}:
\begin{subequations}\label{eq:model_org}
\begin{equation}\label{eq:model_a_org}
    i\hbar\frac{\partial \psi_{1}}{\partial t} = 
    \bigg(-\frac{\hbar^2}{2m_{1}}\frac{\partial^2 }{\partial x^2} +
    u_{1}|\psi_{1}|^2 + u_{12}|\psi_{2}|^2 \bigg)\psi_{1}
\end{equation}
\begin{equation}\label{eq:model_b_org}
    i\hbar\frac{\partial \psi_{2}}{\partial t} =
    \bigg(-\frac{\hbar^2}{2m_{2}}\frac{\partial^2}{\partial x^2} +
    u_{12}|\psi_{1}|^2 \bigg)\psi_{2}
\end{equation}
\end{subequations}
The nonlinear coefficient
$u_{1} = 2 \hbar^2 a_{1}/(m_{1} a^2_{\perp})$ specifies the interaction 
between the superfluid atoms in state $\psi_{1}$ and
$u_{12} =2 \hbar^2 a_{12}/m_{12}(a^2_{\perp} + b^2_{\perp})$  with 
$m^{-1}_{12}=m^{-1}_{1}+m^{-1}_{2}$, determines the coupling 
between the superfluid and impurity atoms. 
Here $a_{1}$ and $a_{12}$ are the characteristic 
scattering lengths while 
$a_{\perp} =\sqrt{\hbar/m_{1}\omega_{\perp}}$ and 
$b_{\perp} =\sqrt{\hbar/m_{2}\omega_{\perp}}$ are the harmonic 
oscillator lengths such that $\omega_{\perp}$ is the 
transverse trapping frequency. The number of atoms 
in each state is conserved by the normalization,
$\int dx |\psi_{i}(x)|^2 = N_{i}$ and $N = N_{1} + N_{2}$
is the total number of atoms. Further, the impurity fraction
$N_{2}/N_{1}$ is kept vey low (within a few percent) in experiments 
and is varied by tuning the rf-amplitude \cite{Aycock_pnas_2017}. 
Accordingly, we neglect the interactions between the impurity atoms.
By means of spatiotemporal scaling \cite{Devassy_pre_2015}, 
$t = \omega^{-1}_{\perp} t^{\prime}, 
x = a_{\perp}x^{\prime},
~\text{and}~ \psi_{i}=\psi^{\prime}_{i}/\sqrt{|a_{\perp}|}$,
Eqs. \eqref{eq:model_a_org} and \eqref{eq:model_b_org} 
can be recast in the following form:
\begin{subequations}\label{eq:model}
\begin{equation}\label{eq:model_a}
    i\frac{\partial \psi_{1}}{\partial t} = 
    \bigg(-\frac{1}{2}\frac{\partial^2 }{\partial x^2} +
    g_{1}|\psi_{1}|^2 + g_{12}|\psi_{2}|^2 \bigg)\psi_{1}
\end{equation}
\begin{equation}\label{eq:model_b}
    i\frac{\partial \psi_{2}}{\partial t} =
    \bigg(-\frac{\rho}{2}\frac{\partial^2}{\partial x^2} +
    g_{12}|\psi_{1}|^2 \bigg)\psi_{2}
\end{equation}
\end{subequations}
where $\rho = m_{1}/m_{2}$ denotes the mass ratio. 
In this study, we consider the mass ratios, $\rho = \{1/2,1,2\}$.
These correspond to the experimentally relevant cases of 
$^{14}\text{K}- ^{87}\text{Rb}$, $^{87}\text{Rb}- ^{87}\text{Rb}$
and $^{87}\text{Rb}- ^{14}\text{K}$ superfluid-impurity systems respectively.
The higher mass ratios of $\rho =3 ~\text{and}~ 4$ apply respectively to the 
$^{133}\text{Cs}- ^{41}\text{K}$ and $^{87}\text{Rb}- ^{23}\text{Li}$ mixtures.
In the resulting dimensionless GP equations the primes 
have been omitted and have the same structure
as above, but with $\hbar =m =1$. The scaled coupling 
coefficients are now defined by, $g_{1} = 2a_{1}/a_{\perp}$ and 
$g_{12} =4a_{12}/a_{\perp}$. 
\par 
In the context of LSA of Eqs. \eqref{eq:model_a} and 
\eqref{eq:model_b}, we examine the MI in a BEC system 
with $n_{10}$ and $n_{20}$ as uniform densities of the 
superfluid and impurity atoms \textit{i.e.}, 
$\psi_{j} = \sqrt{n_{j0}}e^{-i\mu_{j}t}$ for $j =1,2$. 
In terms of the equilibrium densities, chemical potentials, 
$\mu_{1} = g_{1} n_{10} + g_{12}n_{20}$
and $\mu_{2} = g_{12} n_{20}$. 
For the perturbed wave functions of the form, $\psi_{j} = 
(\sqrt{n_{j0}} + \delta\psi_{j})e^{-i\mu_{j}t}$, 
the linearized equations for the perturbations are:
\begin{subequations}\label{eq:pert_eqns}
\begin{equation}\label{eq:pert_eq_a}
\begin{split}
 i\frac{\partial \delta\psi_{1}}{\partial t} = 
    -\frac{1}{2}\frac{\partial^2 \delta\psi_{1}}{\partial x^2} +
    g_{1} n_{10}(\delta\psi_{1} +\delta\psi^{*}_{1}) +%\\
    g_{12}\sqrt{n_{10}n_{20}}(\delta\psi_{2} +
     \delta\psi^{*}_{2})
\end{split}
\end{equation}
\begin{equation}\label{eq:pert_eq_b}
    i\frac{\partial \delta\psi_{2}}{\partial t} =
    -\frac{\rho}{2}\frac{\partial^2 \delta\psi_{2}}{\partial x^2} +
     g_{12}\sqrt{n_{10}n_{20}}(\delta\psi_{1}+ \delta\psi^{*}_{1})
\end{equation}
\end{subequations}
where ``$\ast$" denotes the complex conjugate. 
We consider the perturbations in the form of plane waves,
$\delta\psi_{j} = a_{j} \text{Cos}(kx -\Omega t) + b_{j} \text{Sin}(kx-\Omega t)$
 with real wavenumber $k$, complex eigenfrequency $\Omega$ and $j =1,2$.
Substituting this in  Eqs. \eqref{eq:pert_eq_a} and \eqref{eq:pert_eq_b} 
results in the following dispersion relation for eigenfrequency, $\Omega$:
\begin{equation}\label{eq:dispersion}
%\begin{split}
    \Omega^{4} - \Omega^{2} k^{2}\left(\frac{k^2}{4}\left(\rho^2 +1 \right)+ g_{1}n_{10}\right) +\frac{k^{4}\rho}{4}\left( \frac{k^{4} \rho }{4}+g_{1}n_{10}\rho  - %\\
    4 n_{10}n_{20} g^{2}_{12} \right)= 0
%\end{split}
\end{equation}
Quartic Eq. \eqref{eq:dispersion} for the eigenfrequency, $\Omega$ 
can be solved for its roots and results in:
\begin{equation}\label{eq:eigenvalue}
    \Omega^2_{\pm} = \frac{k^2}{2}\left(\frac{k^{2} \left(\rho^2+1\right)}{4} + g_{1}n_{10}
    \left(1 \pm \sqrt{1+4 \delta g^{2} \delta n -\frac{k^2 \Delta}{2 g_{1}n_{10}}+\frac{k^4 \Delta^2}{16 g^2_{1}n^2_{10}}} \right)\right)
\end{equation}
where $\delta g = g_{12}/g_{1}$ is the interaction ratio, $\delta n = n_{20}/n_{10}$ 
represents the mass imbalance and $\Delta=\rho^{2} -1$. Eq. \eqref{eq:eigenvalue} 
which governs the propagation of perturbations on the top of 
continuous wave solutions may be positive or negative
depending on the nature of interactions and mass 
imbalance.  The continuous wave solutions are stable if 
$\Omega^2_{\pm}>0$ for all real $k$. On the other hand,
the instability gain is defined as $\xi = |\text{Im}(\Omega_{\pm})|$.
In case of scalar BECs without an impurity ($\delta g =\delta n =\Delta =0$),
Eq. \eqref{eq:eigenvalue} reproduces the well-known results 
of MI in dilute BECs where it occurs for an attractive interaction
($g_{1} <0$) in the wavenumber range $0<k<2\sqrt{|g_{1}| n_{10}}$ 
\cite{Agrawal_nlo_2013,Theocharis_pra_2003,Salasnich_prl_2003}. 
The maximum MI gain, $\xi_{\text{max}} = n_{10}|g_{1}|$ 
is attained for $k_{\text{max}} = \sqrt{2n_{10}|g_{1}|}$. 
\par
\begin{figure}[!htb]
    \centering
    \includegraphics[width=0.49\textwidth]{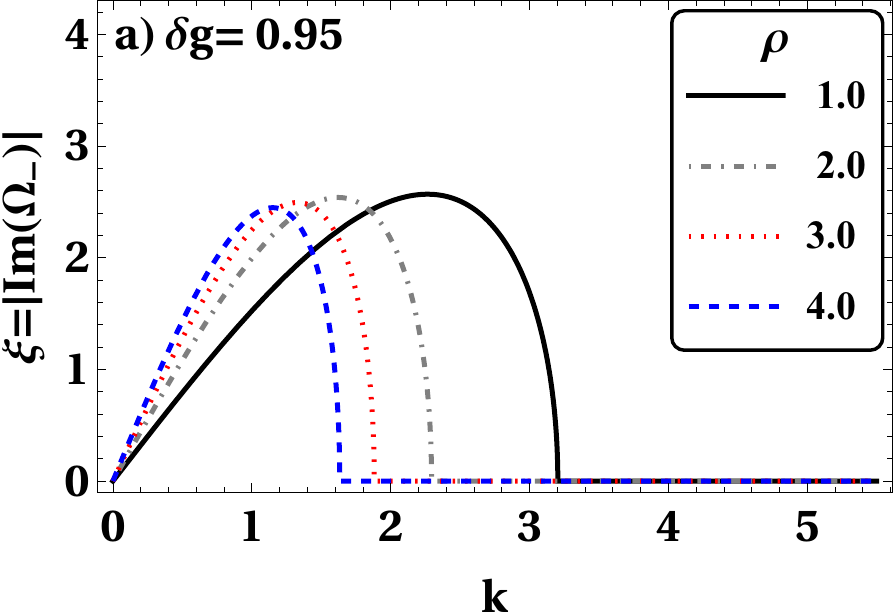}
    \includegraphics[width=0.48\textwidth]{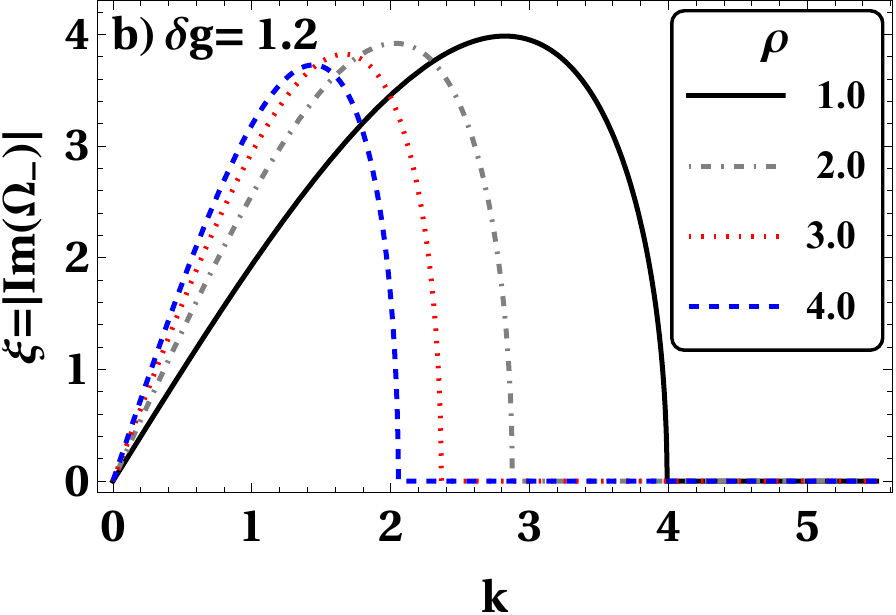}
    \caption{Variation of the MI gain with atomic mass ratio, $\rho$ in a BEC with 
    a) weak ($\delta g <1$) and b) strong ($\delta g >1$) superfluid-impurity interactions,
     The impurity fraction, $n_{20}=0.06$ in each case. }
    \label{fig:gain_rho}
\end{figure}

The effect of the impurities on the MI in BECs can be 
analyzed in the framework of Eq. \eqref{eq:eigenvalue}
whereby it is evident that the BEC is modulationally 
unstable even for $g_{1}>0$. This situation is of peculiar 
importance since the scalar ($\delta g = 0$) \cite{Theocharis_pra_2003} and 
miscible binary BECs ($\delta g <1$) \cite{Kasamatsu_pra_2006} with repulsive $g_{1}$
interactions are modulationally stable.
%%%%%%%%%%%%%%%%%%%%%% Fig 1.
\begin{figure}[!htb]
    \centering
    \includegraphics[width=0.32\textwidth]{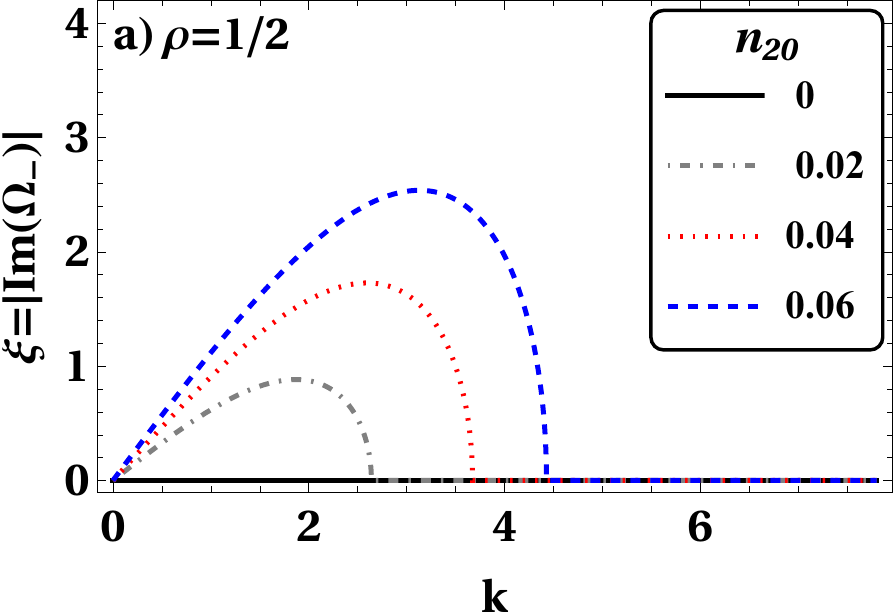}
    \includegraphics[width=0.32\textwidth]{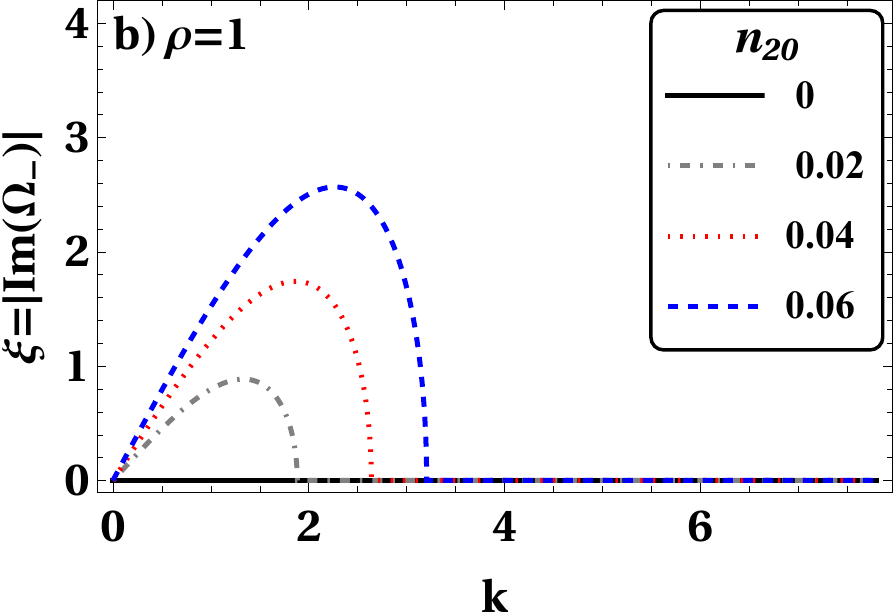}
    \includegraphics[width=0.32\textwidth]{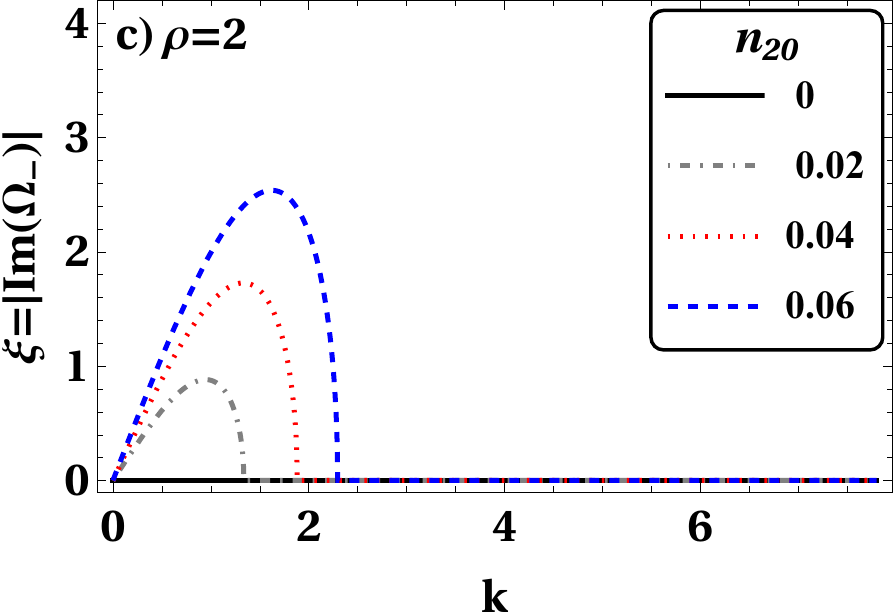}
    \includegraphics[width=0.32\textwidth]{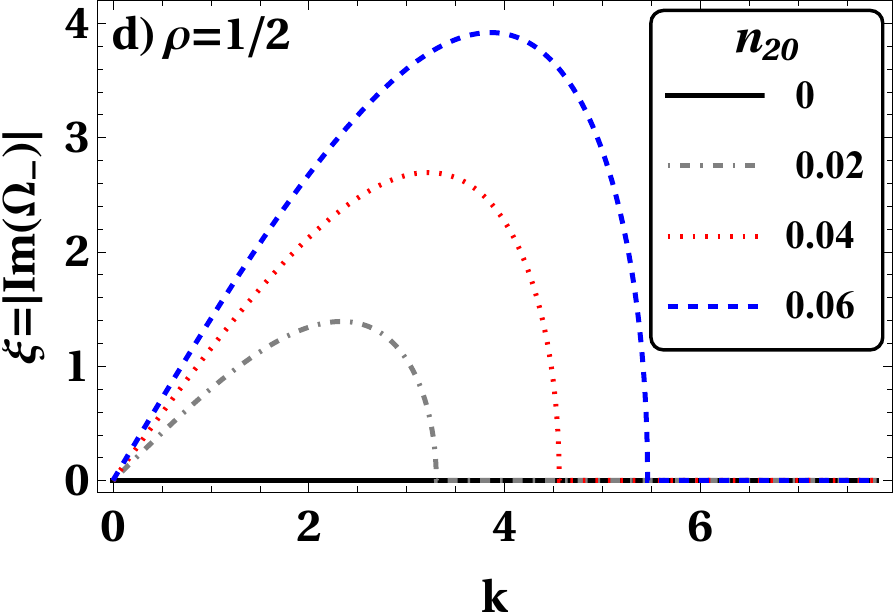}
    \includegraphics[width=0.32\textwidth]{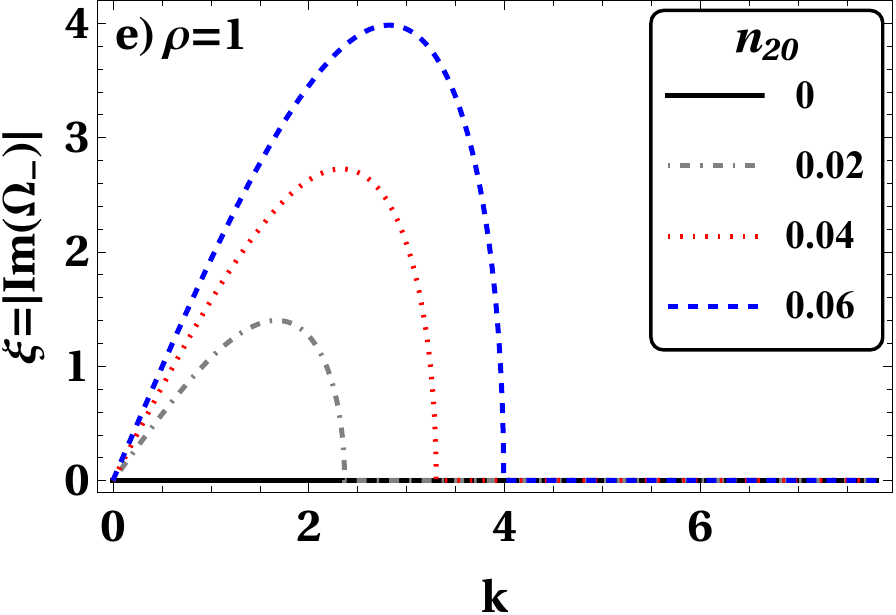}
    \includegraphics[width=0.32\textwidth]{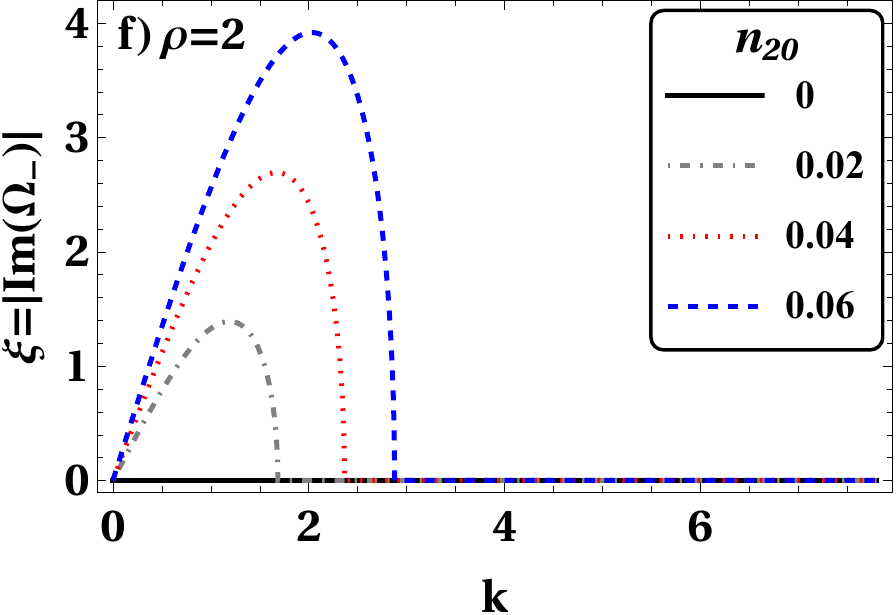}
    \caption{The MI gain corresponding to the $\Omega_{-}$
    perturbation branch for $g_{1} = 50$ and different impurity concentrations, $n_{20}$. The superfluid-impurity coupling is fixed at $\delta g = 0.95$ in (a)-(c) and $\delta g = 1.2$ in (d)-(f) respectively, while $n_{10}+n_{20} =1$. Note that the repulsive scalar BECs with $n_{20}=0$ are always modulationally stable.}\label{fig:gain_1}
\end{figure}
%%%%%%%%%%%%%%%%%%%%%%
A plot of Eq. \eqref{eq:eigenvalue} shown in Figs. 
\ref{fig:gain_1} (a)-(c) display the variation of the MI gain 
with the nature and concentration of the impurities in a repulsive BEC 
with weak superfluid-impurity coupling ($\delta g <1$). 
Here the instability is accounted for by $\Omega_{-}$ 
perturbations only and occurs in the wavenumber range $0<k<
\sqrt{2 g_{1}n_{10}
\left(-1+\sqrt{1+4\delta g^{2} \delta n /\rho}\right)}$. 
 It is evident from Fig. \ref{fig:gain_1}(a)-(c) that 
for a fixed interaction, $\delta g$ and atomic mass, $\rho$ ratios, 
the MI gain and bandwidth increases with the fraction of impurities 
in the BEC. In the simplest case of $\rho =1$, the largest MI gain, 
\begin{equation}\label{eq:gain_max1}
    \xi_{\text{max}}= 
\text{Im}\left(\sqrt{g^2_{1}n^{2}_{10}(-1-2 \delta g^2 \delta n + \sqrt{1+4\delta g^2 \delta n})/2}\right)
\end{equation}
is attained at,
\begin{equation}\label{eq:k_max1}
    k_{\text{max}}= \sqrt{-g_{1} n_{10} + g_{1} n_{10} 
\sqrt{1 + 4 \delta g^{2} \delta n}}.
\end{equation}
 Moreover, the MI gain and the instability bandwidth  decreases 
with the mass ratio, $\rho$ as shown in Fig \ref{fig:gain_rho}.
Consequently, the values of $k_{\text{max}}$ also decrease with $\rho$.
It is worth mentioning that the MI in such BEC systems are also 
independent of the sign of $g_{12}$, though the
tendency of the system against the growth of the 
perturbations increase together with the ratio $\delta g$.
By comparing Figs. \ref{fig:gain_1}(a,b,c) and \ref{fig:gain_1} (d,e,f) 
respectively, it is evident that the MI gain for a given fraction of 
impurities are considerably larger for a stronger 
superfluid-impurity coupling ( $\delta g > 1$).
%The case $\delta g<1$ refers when the impurities are miscible
%within the BEC while $\delta g>1$ refers to the immiscibility
%in the system featuring MI within the multi-component BECs.
\par
\begin{figure}[!htb]
    \centering
    \includegraphics[width=0.32\textwidth]{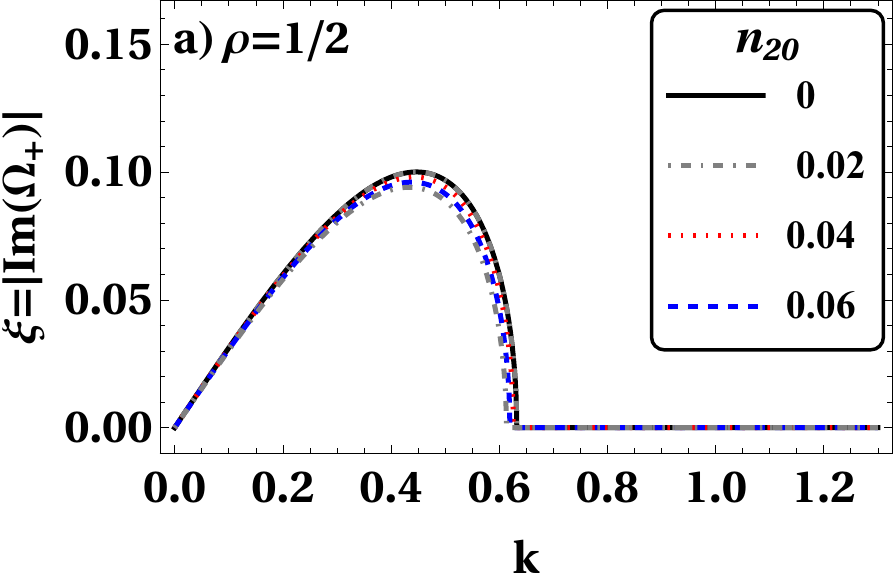}
    \includegraphics[width=0.32\textwidth]{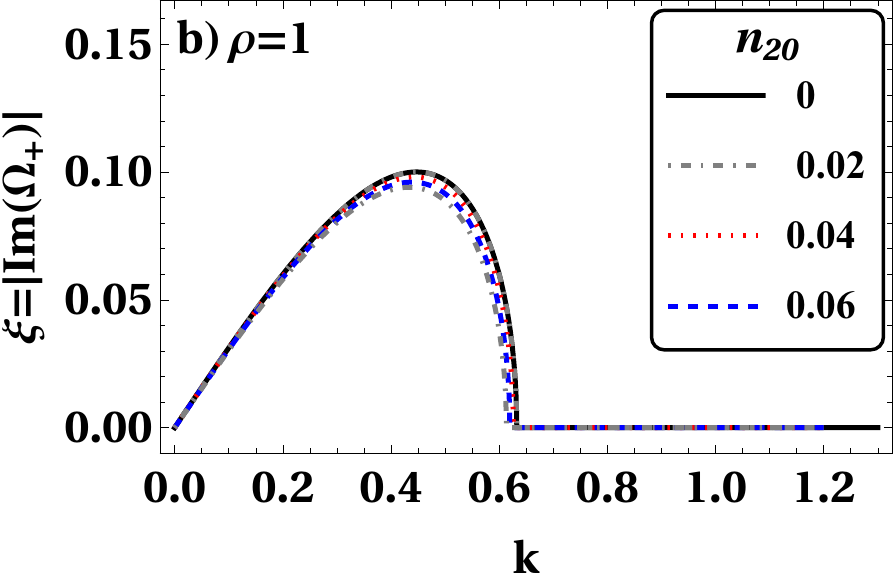}
    \includegraphics[width=0.32\textwidth]{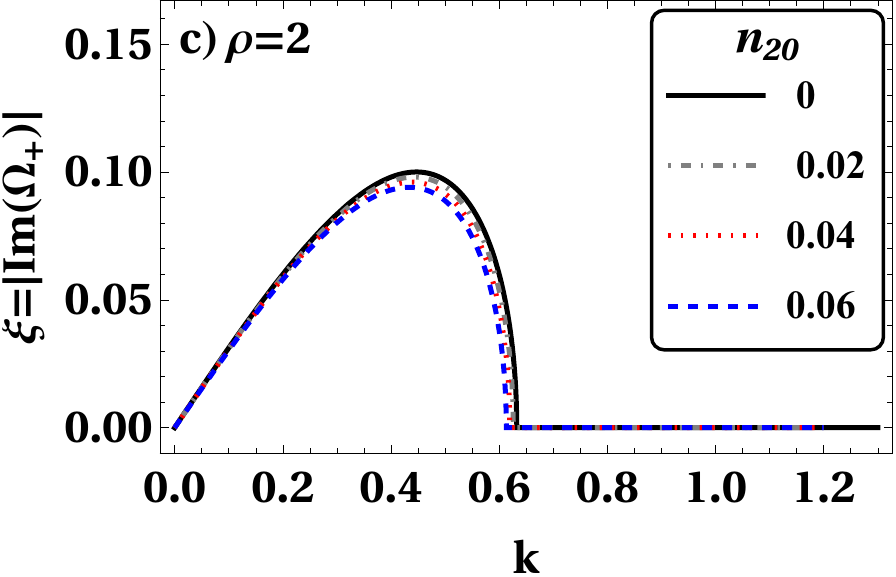}
    \includegraphics[width=0.32\textwidth]{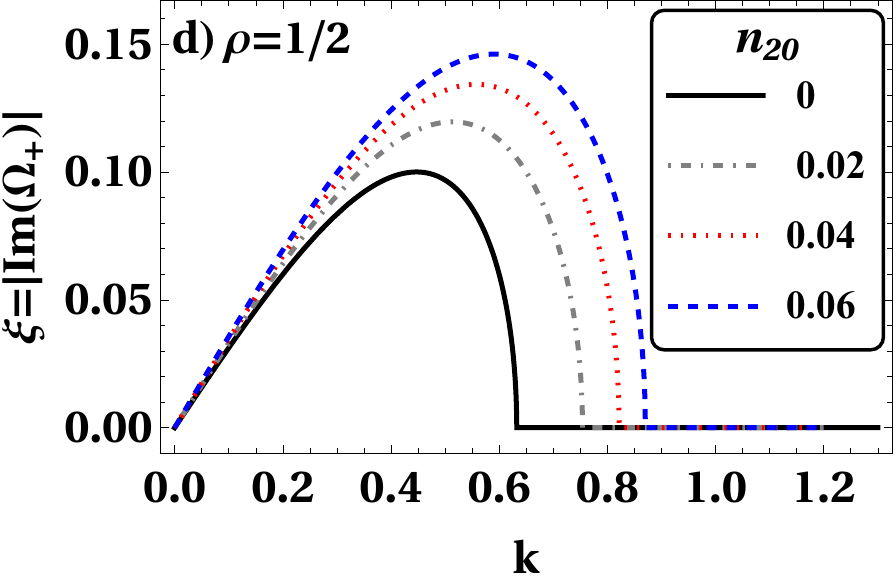}
    \includegraphics[width=0.32\textwidth]{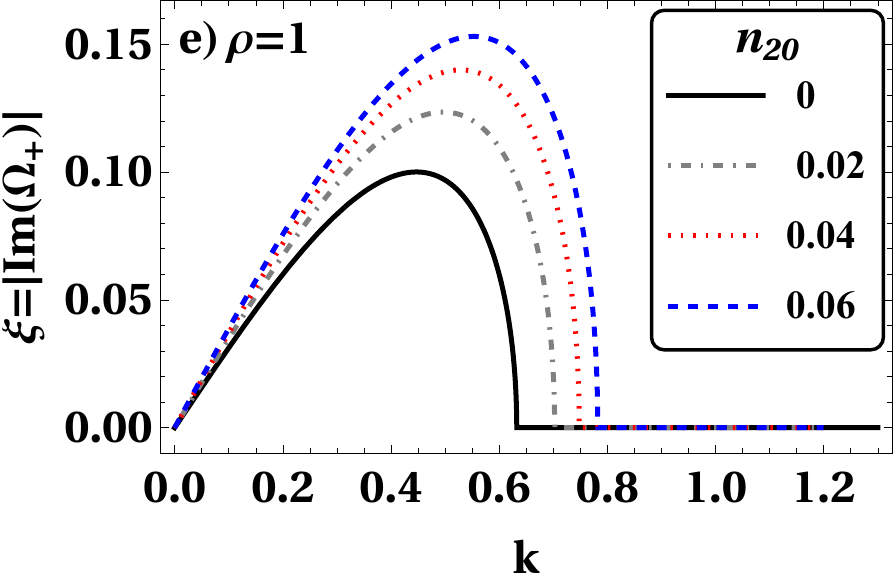}
    \includegraphics[width=0.32\textwidth]{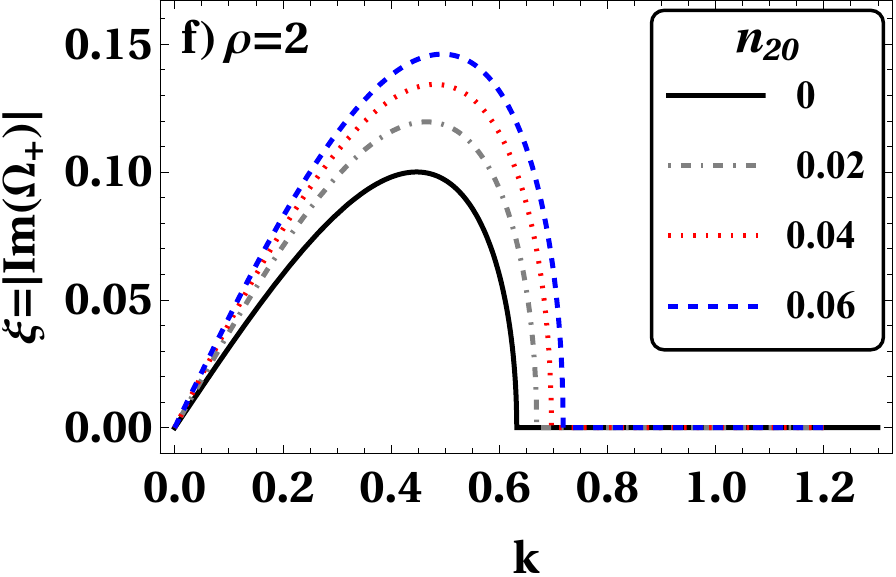}
    \caption{ The MI gain corresponding to the $\Omega_{+}$
    perturbation branch for $g_{1} = -0.1$ and different impurity concentrations, $n_{20}$. The superfluid-impurity coupling is fixed at  $\delta g = 0.05$ in (a)-(c) and $\delta g = 4.0$ in (d)-(f) respectively, while $n_{10}+n_{20} =1$. Note that scalar BECs with  attractive interactions and  $n_{20}=0$ are always modulationally unstable.}\label{fig:gain_2}
\end{figure}

\begin{figure}[!htb]
    \centering
    \includegraphics[width=0.49\textwidth]{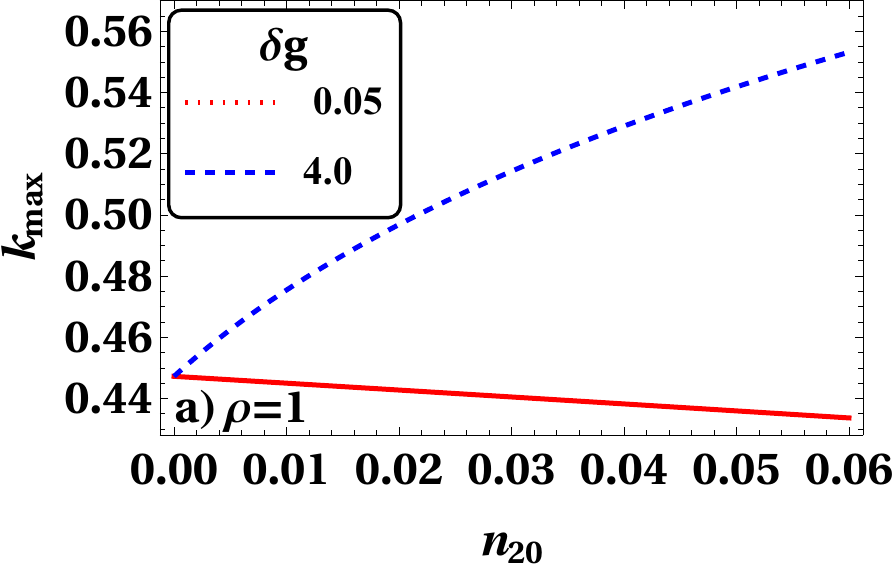}
    \includegraphics[width=0.49\textwidth]{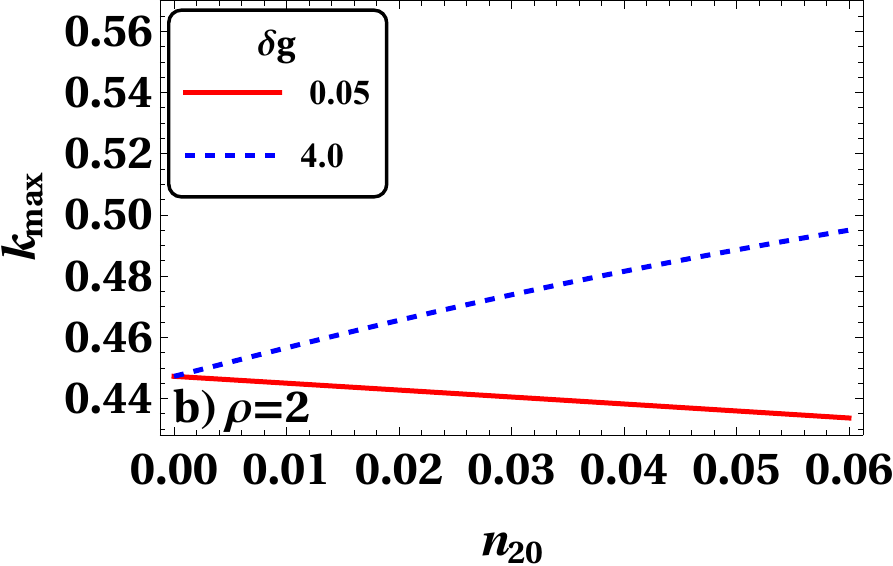}
    \caption{Variation of $k_{\text{max}}$ with 
    the concentration of impurities, $n_{20}$ for 
    $g_{1} = -0.1$, (a) $\rho=1$ and (b) $\rho=2$ .}
    \label{fig:kmax_2}
\end{figure}

 
In an attractive BEC with $g_{1}<0$, the modulational 
instability is accounted for by $\Omega_{+}$ perturbations 
and occurs in the wavenumber range $0<k<
\sqrt{2 g_{1}n_{10}
\left(-1-\sqrt{1+4\delta g^{2} \delta n /\rho}\right)}$
as shown in Fig. \ref{fig:gain_2}. It is evident from 
Fig. \ref{fig:gain_2}(a)-(c) that the MI gain is marginally suppressed with the 
impurity concentration if $\delta g <1$ while for $\delta g >1$, 
the largest MI gain and the instability bandwidth increases 
with increasing impurity fraction in the attractive BECs as shown in 
Fig. \ref{fig:gain_2}(d)-(f). 
%%%%%%%%%%%%%%% Fig. 2
These modifications to the already existing MI
in an attractive BEC due to the impurity fraction can be 
further understood by considering $\rho=1$. 
In such a case, the largest MI gain, 
\begin{equation}\label{eq:gain_max2}
    \xi_{\text{max}}= 
\text{Im}\left(\sqrt{-g^2_{1}n^{2}_{10}(1+2 \delta g^2 \delta n + \sqrt{1+4\delta g^2 \delta n})/2}\right)
\end{equation}
is attained at,
\begin{equation}\label{eq:k_max2}
    k_{\text{max}}= \sqrt{-g_{1} n_{10} - g_{1} n_{10} 
\sqrt{1 + 4 \delta g^{2} \delta n}}.
\end{equation}
A plot of Eq. \eqref{eq:k_max2} in Fig. \ref{fig:kmax_2}(a) shows the variation of 
$k_{\text{max}}$ with the impurity fraction, $n_{20}$ for both $\delta g<1$ and 
$\delta g>1$. Notice that $k_{\text{max}}$ and hence $\xi_{\text{max}}$ 
(not shown in Fig. \ref{fig:kmax_2}) decreases linearly together with 
$n_{20}$ in case $\delta g <1$ while it increases non-linearly 
when $\delta g >1$. The same is true for higher values  of
$\rho$ as shown in Fig. \ref{fig:kmax_2}(b), except that the
$k_{\text{max}}$ values are lower for the corresponding values of 
impurity fraction.
Like in case of BECs with repulsive interactions, 
the MI gain and instability bandwidth decrease 
with $\rho$ for a fixed impurity fraction.
%%%%%%%%%%%%%%%%%%%%%%% Fig 3.
%%%%%%%%%%%%%%%%%%%%%%%% Numerics %%%%%%%%%
\section{Numerics}\label{sec:numerics}
In this section, we discuss the nonlinear dynamics caused by 
impurity-induced MI in the BECs. We Solve 
Eqs. \eqref{eq:model_a} and \eqref{eq:model_b} numerically
by employing FFTW \cite{fftw} mediated split-step fast-Fourier 
method \cite{Pardeep_cpc_2021}. Throughout the simulations,
we set the time step $\Delta t = 0.001$ and the spatial grid 
size $\Delta x = L/N_{s}$ with $N_{s} = 2048$. The domain 
size, $L = 200 ~\text{and}~ 100$ is respectively chosen for 
repulsive ($g_{1}>0$) and attractive ($g_{1}<0$) BECs.
Initial continuous wave background with impurity fraction ranging 
from 0-6\% is studied for its MI by seeding weak and random 
perturbations for different strengths of $g_{1}$ 
and $\delta g$ in the following subsections.  
%%%%%%%%%%%%% subsection a
\subsection{$g_{1} > 0$ and $\delta g <1$}\label{subsec:a} 
\begin{figure}[!htb]
    \centering
    \includegraphics[width=0.8\textwidth]{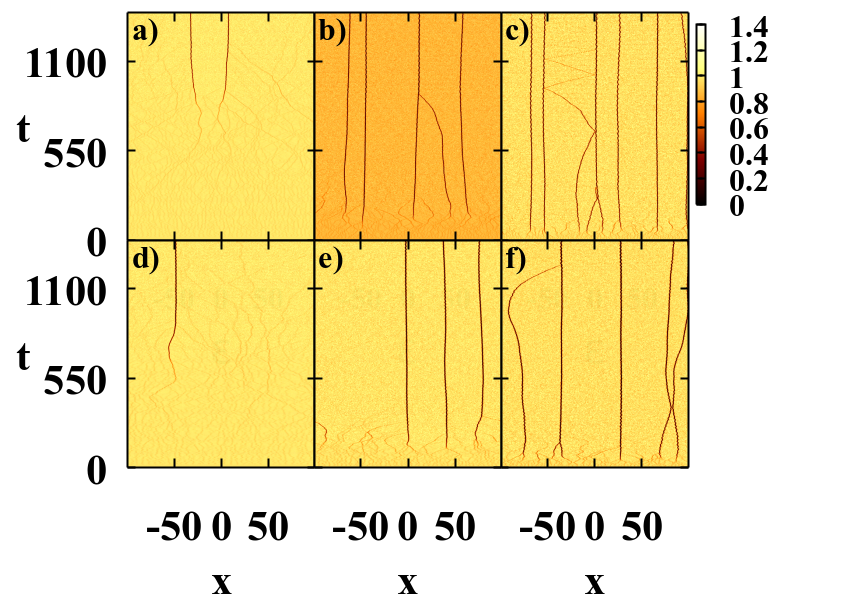}
    \caption{The evolution of the condensate density $|\psi_{1}|^2$ 
    due to the development of the MI by a) 2\%, b) 4\%, and c) 6\%
    impurities respectively in the coordinate space. The upper panel (a)-(c)
    corresponds to $\rho =1$ while the lower panel (d)-(f) represents 
    the corresponding densities with $\rho =2$. The superfluid-impurity
    coupling is maintained at $\delta g =0.05$. The number of 
    solitons increases together with the impurity fraction and decreases with the 
    mass ratio, $\rho$. The dissipation of the generated solitons is negligible 
    for a weaker superfluid-impurity coupling. }
    \label{fig:mi_evol_1}
\end{figure}

The MI analysis discussed in the previous section 
shows that impurities within a BEC make it vulnerable 
towards the growth of perturbations (MI) even for 
repulsive $g_{1}$ interactions and $\delta g < 1$. 
In this regard, Fig. \ref{fig:mi_evol_1}(a)-(c) shows the time
development of the initially perturbed  $\rho =1$  continuous 
wave density for 2\%, 4\% and 6\% impurity concentrations respectively. 
For a given value of $g_{1}=50$ and $\delta g=0.95$, it is evident that 
the BEC realizes phase separation through the formation of dark 
solitary waves. This is because of the repulsive superfluid-impurity 
coupling and the density modulation grows out of phase between the 
two components. This density disturbance in the superfluid increases 
with the increasing concentration of the impurities. Consequently, 
the number of phase-separated domains or solitary waves increases 
proportionately with the concentration of the impurities such that 
for a 2\% impurity, a pair of parallel dark-solitons appear at $t_{\text{MI}}>900$. 
The critical time at which MI-induced solitary waves start 
appearing decreases with the increasing concentration of 
impurities. Though the impurities felicitate the MI 
phenomena in BECs, these are also responsible for 
the dissipation and Brownian motion of the generated 
solitary waves \cite{Aycock_pnas_2017}. 
Fig. \ref{fig:mi_evol_1}(c) shows that one of the 
several generated solitary waves in the superfluid 
executes zig-zag motion during its evolution and 
finally disappears due to the dissipation by impurities.
Since, $\delta g <1$, the dissipation and the Brownian motion of the 
solitary waves due to the impurities is not prominent, 
though it increases with the concentration of the impurities. 
 Similar outcomes are observed in the lower panel 
(Fig. \ref{fig:mi_evol_1}(d)-(f)) where $\rho =2$.
However, in accordance with the analytical results, the number of 
initially generated dark solitons is less than those produced in 
the corresponding  simulations with $\rho =1$.
\subsection{$g_{1} > 0$ and $\delta g >1$}\label{subsec:b}
\begin{figure}[!htb]
    \centering
    \includegraphics[width=0.8\textwidth]{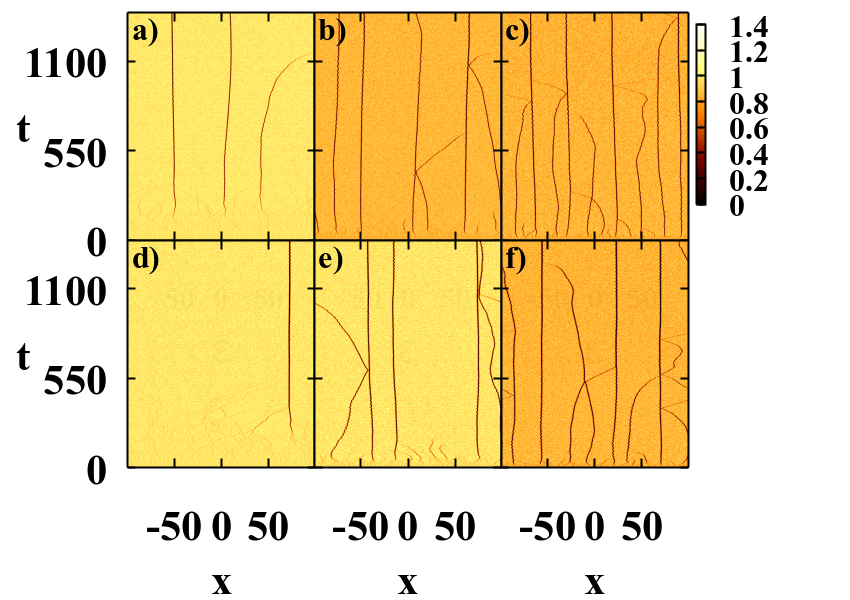}
    \caption{The evolution of the condensate density $|\psi_{1}|^2$ 
    due to the development of the MI by a) 2\%, b) 4\%, and c) 6\%
    impurities respectively in the coordinate space. The upper panel (a)-(c)
    corresponds to $\rho =1$ while the lower panel (d)-(f) represents 
    the corresponding densities with $\rho =2$. The superfluid-impurity
    coupling is maintained at $\delta g =4$. Both the number 
    of solitons and their dissipation are large  for a larger superfluid-impurity 
    coupling.}
    \label{fig:mi_evol_2}
\end{figure}

The condition $\delta g >1$ refers when superfluid-impurity
coupling outbalances the interaction between the superfluid atoms.
From Eq. \eqref{eq:eigenvalue} it is clear that the MI gain increases
with the increase in $\delta g$ for a fixed value of $\rho$ and impurity 
concentration. By comparing the corresponding plots in Fig. \ref{fig:mi_evol_1} 
and Fig. \ref{fig:mi_evol_2}, it is clear that the number of generated  
solitary waves is more for a larger superfluid-impurity coupling 
($\delta g$). Moreover, when the superfluid-impurity interaction is large, 
the dissipation due to the impurity atoms within the BEC is also large. 
Fig. \ref{fig:mi_evol_2} shows that dark solitons within the superfluid 
component dissipate after executing the zig-zag motion. As the number of 
impurities in a BEC increases, more solitons in the superfluid dissipate 
during their time evolution. This consequently reduces the lifetime 
of the generated solitons. Further, by comparing the corresponding plots in
the upper (a-c) and lower (d-f) panels of Fig. \ref{fig:mi_evol_2}, 
it is evident that the number of generated solitons 
decrease with increasing values of mass ratio, $\rho$.
%\FloatBarrier
\subsection{$g_{1} < 0$ and $\delta g <1$}\label{subsec:c}
\begin{figure}[!htb]
    \centering
    \includegraphics[width=0.8\textwidth]{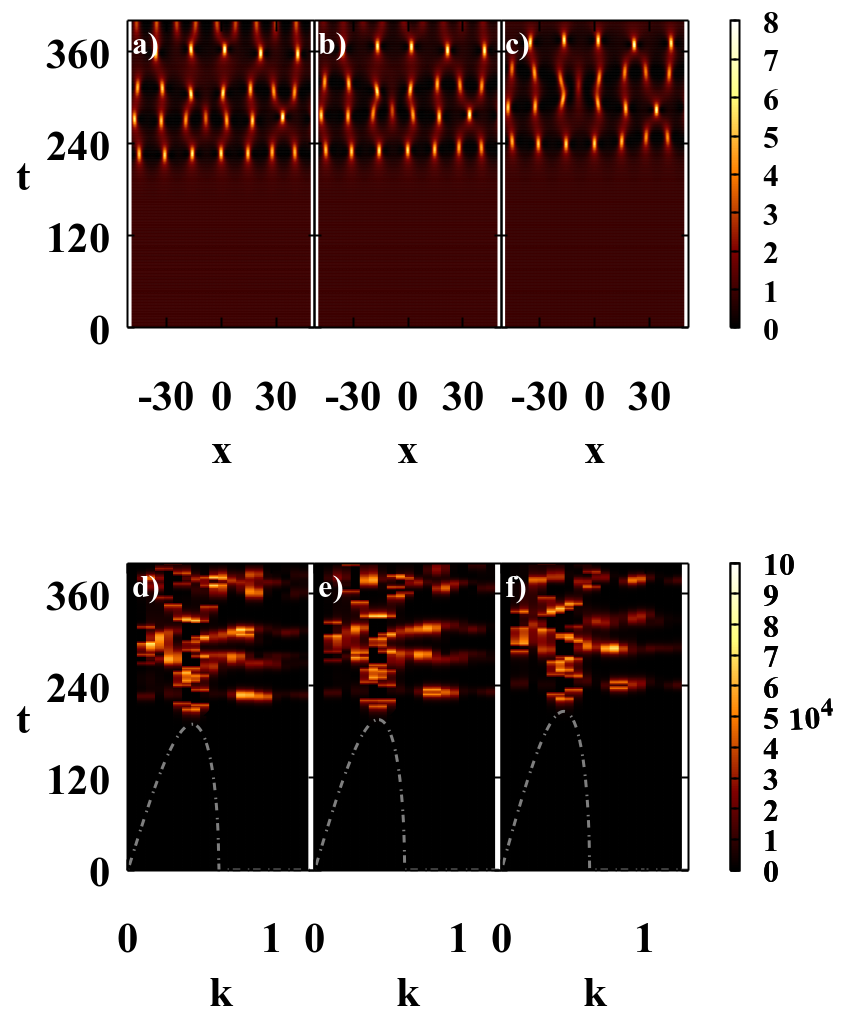}
    \caption{The evolution of the condensate density $|\psi_{1}|^2$ 
    due to the development of the MI by a) 0\%, b) 2\%, and c) 6\%
    impurities respectively in the coordinate space. The lower 
    panel  (d)- (f) represents the corresponding densities in $k$- 
    space. The superfluid-impurity coupling is maintained at 
    $\delta g = 0.95$. For $\delta g <1$,
    the critical time for the generation of bright solitons 
    increases with the impurity fraction.}
    \label{fig:mi_evol_3}
\end{figure}

It is well known that MI in  BECs with attractive 
self-interactions lead to the formation of bright 
matter-wave solitons. Eq. \eqref{eq:eigenvalue} 
and Fig. \ref{fig:gain_2} shows that the BEC 
is modulationally unstable for $g_{1}=-0.1$. However, 
the presence of impurities changes the MI in 
the system only slightly if $\delta g < 1$.
Fig. \ref{fig:mi_evol_3} shows the spatiotemporal
evolution of the initially perturbed continuous wave density 
for different impurity concentrations and parameters, $g = -0.1$, 
$\delta g =0.05$ and $\rho =1$. It shows the generation of a 
train of solitons at $t_{\text{MI}} \ge 205$ in the absence 
of impurities. In accordance with Eq. \eqref{eq:k_max2}, 
the wavenumber corresponding to the initially excited 
mode is $k_{\text{max}} \approx 0.45$ as shown in 
\ref{fig:mi_evol_3}(d). This corresponds to 
$\lambda_{\text{max}} = 2\pi/k_{\text{max}} \approx 14.19$ and 
hence, $n_{\text{s}}=L/\lambda_{\text{max}}\approx 7$ solitons. 
As the concentration of impurities is increased, it is evident
that the critical time, $t_{\text{MI}}$ at 
which MI-induced bright solitons start appearing increases 
slightly due to the slight decrease in the largest MI gain,
$\xi_{\text{max}}$. With an impurity concentration of 
2\% and 6\% respectively, the largest wavenumber, 
$k_{\text{max}} \approx 0.44 ~\text{and}~ 0.43$ correspond 
to the initially excited modes as shown in Figs. 
\ref{fig:mi_evol_3}(e) and \ref{fig:mi_evol_3}(f) respectively. 
However, due to the slight changes in the MI parameters, 
$k_{\text{max}}$ and $\xi_{\text{max}}$, 
the number of initially generated solitons remains the same 
in all cases. Moreover, due to the weak superfluid-impurity 
coupling the dissipation is even negligible.
%%%%%%%%%%%%%%
\subsection{$g_{1} < 0$ and $\delta g >1$}\label{subsec:d}
\begin{figure}[!htb]
    \centering
    \includegraphics[width=0.8\textwidth]{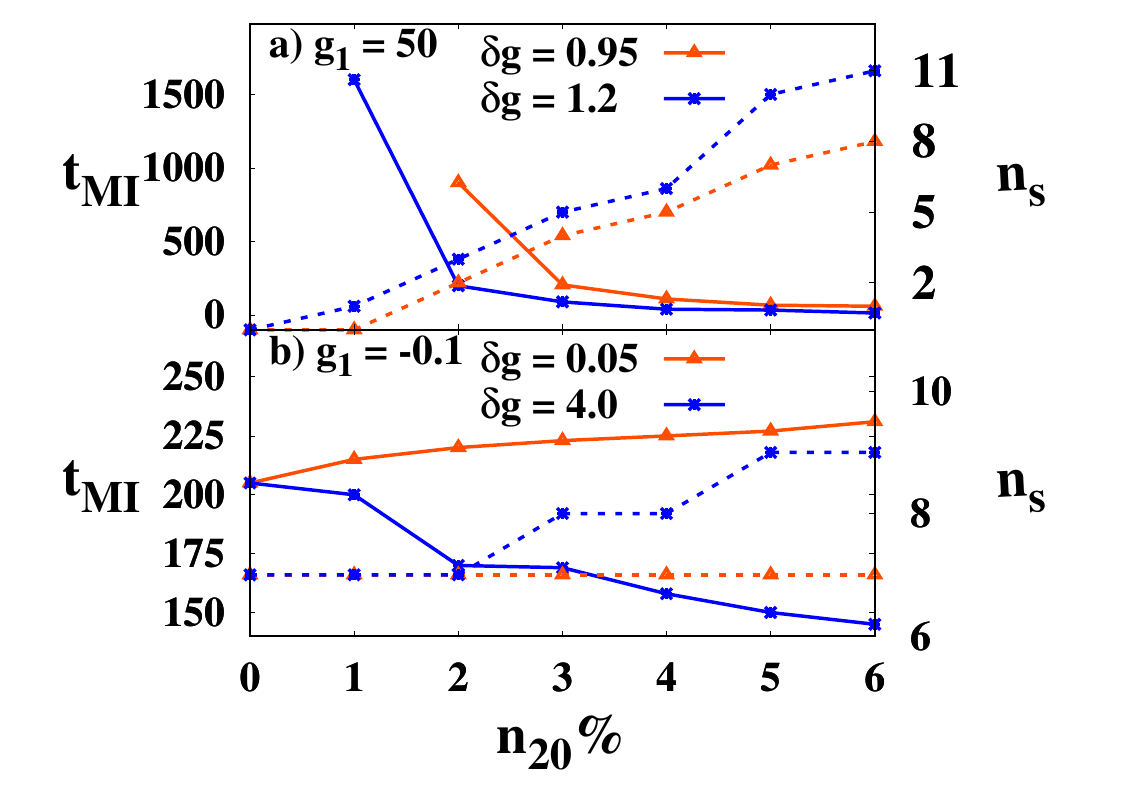}
    \caption{Variation of the MI time, $\text{t}_{\text{MI}}$ (plot-markers with solid lines) and the number of initially generated solitons, ${n}_{s}$ (plot-markers with dashed lines) with impurity concentration, $n_{20}$ in BECs with $\rho =1$.}
    \label{fig:t_mi}
\end{figure}

In the previous subsection \ref{subsec:c}, we showed that the impurities
do not play a significant role in the MI-associated nonlinear dynamics
in attractive BECs for a weak superfluid-impurity coupling 
($\delta g < 1$). The same is true for a larger range of 
$\delta g >1$. In order to explain the effects on the MI 
due to the impurities, we have here chosen a much stronger 
superfluid-impurity coupling, $\delta g = 4$. 
\begin{figure}[!htb]
    \centering
    \includegraphics[width=0.8\textwidth]{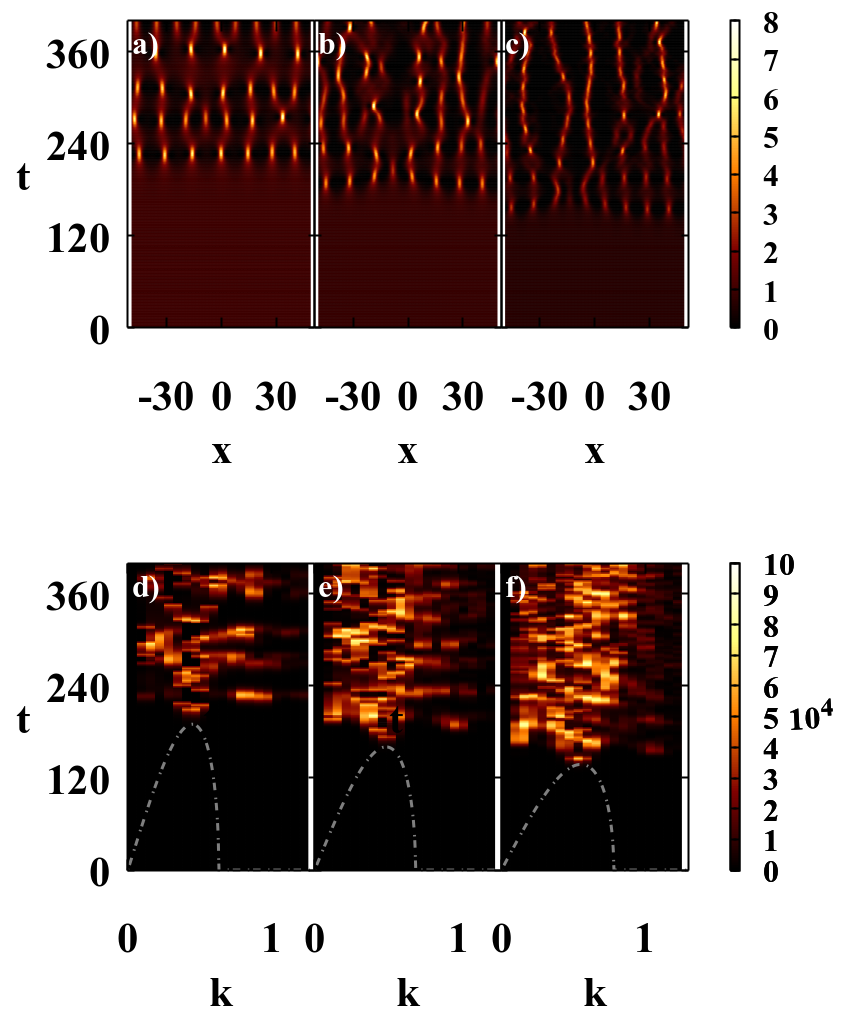}
    \caption{The evolution of the condensate density $|\psi_{1}|^2$ 
    due to the development of the MI by a) 0\%, b) 2\%, and c) 6\%
    impurities respectively in the coordinate space. The lower 
    panel  (d)- (f) represents the corresponding densities in $k$- 
    space. The superfluid-impurity coupling is maintained at 
    $\delta g =1.2$. For $\delta g >1$, the critical 
    time for the generation of bright solitons decreases with the 
    impurity fraction and there is significant dissipation.}
    \label{fig:mi_evol_4}
\end{figure}
Fig. \ref{fig:mi_evol_4} shows the spatiotemporal
evolution of the initially perturbed continuous wave density 
with 0\%, 2\% and 6\% impurity concentrations respectively.
The critical time $t_{\text{MI}}$ decreases 
with the concentration of impurities. 
This is due to the increase in the largest MI gain, 
$\xi_{\text{max}}$ given by Eq. \eqref{eq:gain_max2}. 
With an impurity concentration of 0\%, 
2\% and 6\% respectively, the largest wavenumber, 
$k_{\text{max}} \approx0.45, 0.49 ~\text{and}~ 0.55$  
correspond to the initially excited modes as shown 
in Fig. \ref{fig:mi_evol_4}(d)-(f) respectively. 
Consequently, the number of initially generated solitons,
$n_{\text{s}}=$ 7, 8 and 9 respectively. Moreover, due to 
the strong superfluid-impurity coupling, the dissipation 
of the generated solitons is prominent. This reduces the 
lifetime of the solitons and hence the number of solitons 
decreases with time evolution. The number of initially generated 
solitons and the critical time for MI development as a function of 
impurity concentration and the interaction parameters in the 
numerical simulations for $\rho =1$ is summarized in Fig. \ref{fig:t_mi}.
The MI time always decreases with the impurity concentration and 
the strength of superfluid-impurity coupling in BECs with repulsive 
$g_{1}$ nonlinearity while in attractive BECs, the MI time either 
decreases or increases with the impurity fraction, depending on 
the strength of the superfluid-impurity coupling. In the latter case 
with $\delta g <1$, the MI time increases  while if $\delta g >1$ 
MI time decreases.
%%%%%%%%%%%%%%%%%%%%%% Table %%%%%%%%%%%%%%%%%%%%%%%%%%%%%%%%%%%%%%%
%\begin{center}
%\vspace{-0.8 cm}
%\begin{table}[!thb]
%\caption{The summary of the MI parameters characterizing the 
%spatiotemporal evolution of a BEC in presence of impurities, $n_{20}$ such that $g_{1}$ %is the interaction between the superfluid atoms,
%$\delta g$ specifies the superfluid-impurity coupling and %$n_{10}+n_{20}=1$.}\label{tab1} %\vspace*{-4cm}
%{\footnotesize \centering
%\begin{tabular}{p{1.0cm}p{1.0cm}p{1.3cm}p{1.3cm}p{2.6cm}c}
%\hline\hline
%$g_{1}$ & $\delta g$ & $n_{20}$ (\%) & $t_{\text{MI}}$ & Number of solitons &\\[0.5ex] \hline
%\multirow{7}{*}{$ 50$}
%& $0$ & $0$ & $\infty$ & $ 0$ (stable) &\\
%& \multirow{3}{*}{$0.95$} & $2$ & $900$ & $ 2$ &\\
%                        & & $4$ & $110$ & $ 4$ &\\
%                        & & $6$ & $60$ & $ 6$ &\\
%& \multirow{3}{*}{$1.2$} & $2$ & $200$ & $ 3$ &\\
%                       & & $4$ & $40$ & $ 4$ &\\
%                       & & $6$ & $15$ & $ 6$ &\\
%\hline
%\multirow{5}{*}{$-0.1$}
%& $0$ & $0$ & $205$ & $ 7$ (unstable)&\\
%& \multirow{2}{*}{$0.05$} & $2$ & $210$ & $ 7$ &\\
%                        & & $6$ & $225$ & $ 7$ &\\
%& \multirow{2}{*}{$4$}    & $2$ & $170$ & $ 8$ &\\
%                        & & $6$ & $140$ & $ 9$ &\\
%\hline\hline
%\end{tabular}
%}
%\end{table}
%\end{center}
%%%%%%%%%%%%%%%%%%%%%%%%%%%%%%%%%%%%%%%%%%%%%%%%%%%%%%%%%
\section{Conclusion}\label{sec:conc}
In summary, we investigated the MI by employing linear stability 
analysis and direct numerical simulations in a BEC coupled with a dilute 
fraction of Bose-condensed impurities. Being dilute, 
the coupling among the impurity atoms is neglected. It is well established 
that single-component BECs  with repulsive interactions are always 
modulationally stable. Moreover, the binary BECs with an equal 
proportion of the two components are modulationally unstable 
iff the cross-phase mediated repulsion between the components 
outbalances their self-repulsion. However, in the presence of 
dilute impurities, BECs are always modulationally unstable and 
is independent of the sign of the superfluid-impurity interaction.
The MI induces spatial pattern formation in a repulsive 
BEC and the generated domains have a solitary wave structure.
The tendency of the BEC towards MI increases with the 
increasing impurity fraction for a fixed superfluid-impurity 
coupling strength. Consequently, the number of domains increases
together with the fraction of impurities in the BEC.  Moreover, 
the MI gain in a given BEC, the instability  bandwidth, the number 
of solitons as well as $k_{\rm max}$, the value of the wave vector 
$k$ at which the MI gain attains  maximum, decreases with the 
decreasing mass of the impurity atoms. Despite increasing 
the tendency of a repulsive BEC towards MI, the impurities 
are also responsible for the Brownian motion of the solitons
and their dissipation. This reduces the lifetime of the solitary
wave structures. Naturally, the dissipation is prominent for a 
greater impurity fraction and increasing superfluid-impurity 
coupling strength. 
\par 
As concerns the presence of impurities in attractive 
BECs slightly affect the preexisting MI phenomena for a weak 
superfluid-impurity coupling ($\delta g <1$). The modulational 
instability time only slightly increases with the impurity fraction. 
However, for a strong superfluid-impurity coupling 
($\delta g >1$) the modulational instability time decreases 
and the number of generated solitons increases with the 
impurity fraction. 
\section{Acknowledgements}
I A Bhat acknowledges CSIR, 
Government of India, for funding via CSIR 
Research Associateship (09/137(0627)/2020 EMR-I). 
BD thanks Science and Engineering Research
Board, Government of India for funding through research 
project CRG/2020/003787.
%%%%%%%%%%%%%%%%%%%%%%%%%%%
%%
\bibliographystyle{elsarticle-num} 
\bibliography{references}

\end{document}